# Polymer Reptation and Nucleosome Repositioning

H. Schiessel[1,2,*], J. Widom[3], R. F. Bruinsma[1,†] and W. M. Gelbart[2]

[1]Department of Physics, and [2]Department of Chemistry and Biochemistry, University of California, Los Angeles CA 90095-1569, and [3]Department of Biochemistry, Molecular Biology, and Cell Biology, Northwestern University, Evanston IL 60208

ABSTRACT

We consider how beads can diffuse along a chain that wraps them, without becoming displaced from the chain; our proposed mechanism is analogous to the reptation of "stored length" in more familiar situations of polymer dynamics. The problem arises in the case of globular aggregates of proteins (histones) that are wound by DNA in the chromosomes of plants and animals; these beads (nucleosomes) are multiply wrapped and yet are able to reposition themselves over long distances, *while remaining bound by the DNA chain*.

\* Present address: Max-Planck-Institute for Polymer Research, Theory Group, POBox 3148, 55021 Mainz, Germany
† Present address: Instituut-Lorentz for Theoretical Physics, Universiteit Leiden, Postbus 9506, 2300 RA Leiden, The Netherlands

The organization of DNA in chromosomes is still poorly understood except on the level of *primary* structure. In particular, it has been known for twenty-five years now that the basic unit of chromatin – the nucleosome – consists of DNA wrapped almost two times around a globular octamer of cationic proteins (histones) about 100Å in diameter; typical "linker" lengths of DNA chain in between successive nucleosomes are also of order 100Å [1]. The higher-order "folding" of this beads-on-a-chain motif, into successive secondary and tertiary structures on scales from 100's of Ångstroms to microns, is yet to be elucidated, even though it is clearly of fundamental importance for understanding a host of biological processes ranging from gene expression to cell division. While the individual nucleosome structures are now documented in exquisite detail, largely from high-resolution X-ray analyses [2], relatively little is known about their *dynamics*. Many studies have established that RNA polymerases can act "through" a nucleosome [3], in the sense that transcription occurs involving histone-wrapped DNA sequences. Controversy remains concerning the extent to which DNA is displaced from the histones as the polymerase acts, e.g., does it form intra-nucleosomal loops or does it dissociate completely [4]?

Recent *in vitro* experiments have addressed a related question which involves a simpler aspect of nucleosome dynamics. Specifically, polymerases (and all other proteins other than the histones comprising the core octamer) are absent and one studies directly the mobility of a "naked" nucleosome along the DNA chain that wraps it. One takes advantage of the fact that different nucleosome positions in the chain give rise to different electrophoretic mobilities [5], and that the motion of the nucleosome along the chain can be surpessed by subphysiological temperatures or ionic strengths, and by the presence of $Mg^{2+}$ [6]. Nucleosomes are prepared on DNAs that are a few hundred basepairs (bps) in length (a few nucleosome lengths). The (apparently) equilibrium distribution of positioning isomers that is initially created is separated by an initial dimension of gel electrophoresis (nucleosomes nearer the ends migrate faster than more centrally located ones [6]) in conditions where mobility is suppressed. An entire track from such a gel is excised, incubated for some period of time in new conditions where mobility may occur, then changed back to conditions where mobility is suppressed again, and run in a second, equivalent, dimension of gel electrophoresis. Essentially, the first dimension of electrophoresis creates a nonequilibrium distribution; this may relax during the subsequent incubation, which in turn will be manifest as product moving off the diagonal in the second dimension of electrophoresis. In this way one time resolves the motion of nucleosomes from an initial site to a set of final positions, and is able to demonstrate that *these motions occur without the nucleosome becoming unbound from the DNA chain* [6]. Furthermore, it is significant that the *in vitro* repositioning is a passive process involving nothing more than equilibrium fluctuations, as opposed to the ATP-driven enzymatic processes involved *in vivo* [7].

In this Letter we argue that the repositioning of nucleosomes without dissociation from the DNA chain that wraps them can be understood through the diffusional motion of intra-nucleosomal *loops*. This *biological* process is closely analogous to the now-familiar *physical* situation of reptation of "stored length" in polymer chains. Almost thirty years ago de Gennes [8] discussed the motion of a flexible chain trapped in a gel, modeled by a matrix of point, fixed, obstacles that could not be crossed by the polymer. Fig. 1 depicts schematically the mechanism whereby diffusion of these "defects" of stored length (*b*) gives rise to overall translation of the chain; specifically, when the loop moves through the monomer at B, this monomer is displaced by a distance *b*. de Gennes wrote down a conservation equation for this motion of defects along the trapped chain and calculated its overall mobility, and thereby, in particular, the molecular-weight dependence of the overall translation diffusion coefficient. In our present situation the reptation dynamics do not arise from obstacles due to a host matrix (as in a gel) or to other chains (as in a melt), but rather to loops associated with unsaturated adsorption of the DNA on the protein complex. Similar physics appears to arise in the lateral displacements of a linear polymer adsorbed on a *bulk* solid surface. Granick et al. [9], for example, have measured the translational motion of adsorbed PEO on functionalized (hydrophobic) silica, specifically the dependence of its center-of-mass diffusion constant ($D$) on molecular weight ($M$). They find an unusual scaling behavior – $D \propto M^{-3/2}$ – but one that is accounted for by "slack between sticking points", so that lateral motion of the polymer proceeds via a caterpillar-like diffusion of chain loops. In the case of intra-nucleosomal loops considered in the present paper, we are essentially in the limit of infinite molecular weight, because of the chain length being large compared to the bead (solid substrate) diameter. Furthermore, we deal with a lower dimensional problem, since our chain is wrapped (absorbed) on a 1D path rather than a 2D surface. But the basic features of loop formation and diffusion, and subsequent motion of the overall chain – in particular the exclusive role of equilibrium fluctuations in driving these processes – are the same in both cases.

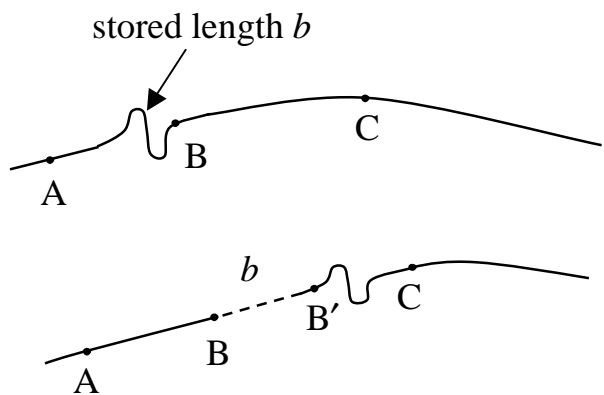

Fig. 1. *The de Gennes-Edwards mechanism of translational motion of trapped chains via diffusion of "stored length defects"; the monomer at B moves by an amount b when the loop passes through it.*

As shown in earlier studies of competitive protein binding to nucleosomal DNA [10], thermal fluctuations lead to lengths of the chain becoming unwrapped at the ends of its adsorbed portion. If some length of linker is pulled in before the chain re-adsorbs, then an intra-nucleosomal loop is formed – see Fig. 2. As mentioned above, such loops have been implicated in recently proposed mechanisms for transcription through nucleosomes by RNA polymerase, but there the loops need to be large enough to accommodate the bulky enzyme whereas this requirement does not arise in our case of "naked" nucleosomes. We calculate first the equilibrium shape and length distribution of these loops, in terms of the chain bending elasticity ($\kappa$), adsorption energy per unit length ($\lambda$), and protein aggregate size ($R_0$). We then consider the diffusion of these loops from one end of the nucleosome to the other. Finally, treating this motion as the elementary step in the diffusion of the nucleosome itself along the wrapping chain, we are able to make estimates of the nucleosome repositioning rates as a function of $\kappa$, $\lambda$, $R_0$, and viscosity $\eta$. These estimates are consistent with measured rates [5,6] and suggest several new experiments for testing this mechanism of repositioning. Finally, we speculate on the possible connections of these dynamics to the motion of motor proteins such as the transcriptionally-active RNA polymerases.

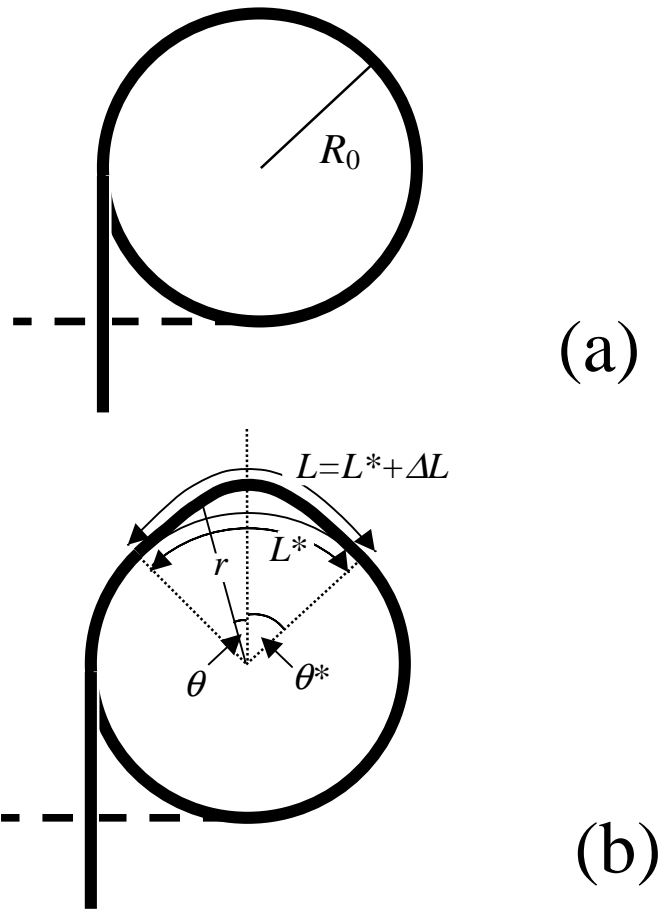

Fig. 2 (a) *Top view, looking down the histone octamer's cylinder axis, showing the DNA chain beginning its wrapping of the cylinder's surface; the remaining 3/4 turn lies below the chain shown. (b) Loop formation involving length $\Delta L$ of linker chain being incorporated into the nucleosome, with length $L^*$ of "exposed" surface. r and $\theta$ are the polar coordinates of an arbitrary position on the loop; $+\theta$ and $-\theta$ correspond to the points were the loop joins the surface.*

We start by calculating the equilibrium statistical mechanical probability associated with the formation of an intra-nucleosomal loop of arbitrary size. When no loops are present, a nucleosome consists of a length $L_0$ of DNA (chain) wrapped continuously around the histone octamer (ball), see Fig. 2a. In reality, the configuration of the adsorbed chain is a superhelix (of contour length $L_0$) spanning the full height of a right-circular-cylinder. The length (147 bp's [2]) and shape of the wrapped anionic chain is determined by the particular distribution of cationic charges on the octamer surface. Furthermore, the strongest adsorption – "sticking" – sites correspond to those points where the minor groove of the DNA is facing inward toward the protein surface; accordingly they are separated by the helical pitch ($\cong 10$ bps). We can proceed, however, without making any explicit assumptions about the shapes of *either* the histone octamer *or* the wrapped DNA.

Consider a fluctuation in which some length of the chain becomes unwrapped (this can only happen at the *end* of the adsorbed portion of chain) *and* simultaneously some length, say $\Delta L$, of linker (i.e., previously *un*adsorbed chain) is "pulled in" before the chain *re*-adsorbs. Upon re-adsorption we are back to where we started if $\Delta L = 0$; but for *non*vanishing $\Delta L$ the fluctuation results in a loop of contour length $L*+\Delta L$ being formed; here $L*$ is the "exposed" length of nucleosome associated with the loop; see Fig. 2b. We shall consider $\Delta L$ values that are multiples of 10bp lengths, since this is the periodicity of the DNA helical pitch (see above). For the energy associated with forming a loop of this kind, we can write

$$\Delta U = \frac{\kappa}{2} \int_0^{L*+\Delta L} \frac{ds}{R^2(s)} + \left(\lambda - \frac{\kappa}{2R_0^2}\right) L*, \qquad (1)$$

where $\kappa = l_P k_B T$ is the 1D bending elastic constant of the semiflexible chain (persistence length $l_P$), $\lambda$ is the adsorption energy per unit length, $R_0$ is the radius of the ball, and $1/R(s)$ is the local curvature of the loop at distance *s* along its contour.

Since the configurations of small loops can be assumed to be planar, it is convenient to describe them in terms of the function $r(\theta)$ – cf. Fig. 2b – where *r* and $\theta$ are the polar coordinates of an arbitrary point on the loop (with the origin chosen at the ball center, with the *X*-axis running through the center of the loop, assumed asymmetric). The loop curvature is given by $1/R = |r''r - 2r'^2 - r^2|/(r^2 + r'^2)^{3/2}$, with primes and double-primes denoting the first and second derivatives, respectively, with respect to $\theta$. Upon writing $r(\theta) = R_0 + u(\theta)$ and keeping only through quadratic terms in *u* and its derivatives (*u* is assumed to be small compared to $R_0$ [11]), $\Delta U$ follows as an explicit functional of *u*. We then minimize this functional with respect to $u(\theta)$, subject to the constraint that the loop contour length (another functional of *u*) is $L*+\Delta L$. This leads to an Euler-Lagrange equation of the form

$$u^{(4)} + \left(\frac{5}{2} + \frac{TR_0^2}{\kappa}\right) u^{(2)} + u = 0. \qquad (2)$$

Here $u^{(n)}$ denotes the *n*th derivative of *u* with respect to $\theta$, and *T* is the Lagrange multiplier that constrains an extra length $\Delta L$ to be adsorbed as the loop is formed. The relevant boundary conditions are that *u* be symmetric about $\theta = 0$ and that *u* and its first derivative vanish at the "boundaries", i.e., at $\theta = \theta* \; (= L*/2R_0)$ and $\theta = -\theta^*$.

Solving the corresponding Euler-Lagrange equation subject to the above constraint and boundary conditions gives us the *trajectory* $u(\theta)$ of the loop for a given $L^*$ and $\Delta L$. Substituting this result into $\Delta U$, and minimizing with respect to $L^*$, gives

$$L^* \cong \left(\frac{20\pi^4 \kappa}{\lambda R_0^2}\right)^{1/6} \left(\frac{\Delta L}{R_0}\right)^{1/3} R_0, \qquad (3a)$$

from which it follows in turn that

$$\Delta U \cong \varepsilon \left(\frac{\Delta L}{R_0}\right)^{1/3} = \left[\frac{20\pi^4 \kappa}{R_0}(\lambda R_0)^5\right]^{1/6} \left(\frac{\Delta L}{R_0}\right)^{1/3}. \qquad (3b)$$

The probability distribution for formation of loops of size $\Delta L$, then, is simply given by the corresponding Boltzmann factor, normalized so that the number of loops for $\Delta U$ vanishing would be the maximum that is geometrically possible, i.e., $L_0/L^*$:

$$n_{eq}(\Delta L) \cong \frac{L_0}{L^*} \exp(-\Delta U/k_B T). \qquad (4)$$

(Note that the prefactor accounts for the entropy of the loop positions).

Now, what about the loop *dynamics*? The key idea here is that diffusion of the histone octamer along the DNA is achieved by formation and annihilation of loops. Let $D$ denote the diffusion constant relevant to this motion of the ball along the chain, and let $B = 1/T$ be the rate at which loops are formed (by incorporation of linker length); $D = B(\Delta L)^2$. These loops "disappear" due to their diffusion "off" the ball. As we will show *a posteriori* the lifetime $t = 1/A$ of a loop is much shorter than the average time $T$ required to form one loop, i.e., $t \ll T$. Thus the average number of loops is given by $t/T = B/A$. At equilibrium this number is given by the Boltzmann expression for $n_{eq}$ [see (4)]. Since this number will be shown below to be much smaller than unity, we are justified in assuming that only one loop at a time needs to be considered in treating the diffusion of intra-nucleosomal loops. Hence $B$, the rate of loop formation, is given by $B = An_{eq} \cong A(L_0/L^*)\exp(-\Delta U/k_B T)$, and $D$ by $B(\Delta L)^2$.

It remains only to evaluate $A$, characterizing the rate of diffusion of loops "off" a ball. Let $D^+$ denote the diffusion constant associated with this motion. [$D^+$ characterizes the diffusion of *loops* through a wrapped ball, as opposed to the coefficient $D$ (see above) that describes diffusion of the *ball* along the chain.] Since the distance which the loop must move to leave the ball is $L_0$, the total length of adsorbed chain, we can write $A^{-1} \cong L_0^2/D^+$. From the Stokes-Einstein relation we have furthermore that $D^+ = k_B T/\zeta$ where $\zeta \cong \eta L^*$ is the friction coefficient, with $\eta$ the effective solution viscosity. $L^*$, as before, is the exposed length of ball associated with the loop, and hence provides the loop size relevant to its diffusion along the (1D!) nucleosome path of the chain. Our estimation of $\zeta$ by $\zeta \cong \eta L^*$ is based on two assumption: (*i*) The loop diffusion is governed by its Brownian motion through the fluid and not by the unbinding/rebinding events of the sticking sites. (*ii*) A loop (with a given stored length $\Delta L$) has a well-defined size $L^*$. The hydrodynamic analyis (i.e. assumption (*i*)) is justified because each loop diffusion step requires unbinding of only a single sticking site, whose binding energy is of order $k_B T$ (see below). Even though this binding energy is rather weak the concept of a loop with a well-defined size $L*$ is justified. It can be shown (see below) that the lenght of a loop on a nucleosome shows only small deviations from its optimal value. Combining all of the results from this and the preceding paragraph then gives

$$D \cong \frac{k_B T}{\eta L_0}\left(\frac{\Delta L}{L*}\right)^2 \exp(-\Delta U/k_B T). \qquad (5)$$

with $L*$ given by Eq. (3a).

Recalling our earlier expressions [cf. (3a,b)] for $L*(\Delta L)$ and $\Delta U(\Delta L)$, and taking reasonable numerical estimates for $\eta$ (a centipoise), $R_0$ (50Å), $L_0$ (500Å), $\kappa$ (500$k_B T$Å) [12], and $\lambda$ ($(1/20)k_B T$/Å), i.e., $2k_B T$ per sticking site [1,10], we find that $D$ is of order $10^{-12}\, cm^2/s$ for $\Delta L = 10$ base pairs (34Å). The formation energy $\Delta U$ associated with this minimum loop size is about $7 k_B T$, and the corresponding number of loops per ball is negligible compared to one, vindicating our neglect of loop-loop interactions. Furthermore it follows from Eq. (1) that the energy $\Delta U$ of a loop as function of its size $L*$ (for $\Delta L = 10$ basepairs) is given by $\Delta U(L*)/k_B T \approx L*/(20\text{Å}) + 56 \times 10^{10}\, \text{Å}^5/(L*)^5$. From the distribution of loop sizes $p(L*) \propto \exp(-\Delta U(L*)/k_B T)$ it follows that a loop has typical size fluctuations of the order of 30Å around its optimal value $L* \approx 200$Å. Thus *both* assumptions discussed above Eq. (5) are valid at the same time: The opening of a binding site (a change of $L*$ by 36 Å) requires only an energy of the order $k_B T$ *and* the loop size distribution is sharply peaked around $L*$. Note that diffusion coefficients of order $10^{-12}\, cm^2/s$ imply times of order *seconds* for histone octamers to reposition themselves over distances of order tens of base pairs. For $\Delta L$ twice [three times] its minimum value, our $D$ is seven [thirty] times *smaller*, etc. Thus it appears that the dominant fluctuations driving the formation and diffusion of intra-nucleosomal loops (and hence the diffusion of the nucleosome along the DNA) are those involving a minimum length (10 base pairs, 34Å) of linker being absorbed.

It is not possible at present to compare our results more quantitatively with experiment, since the measurements performed to date all involve one hour rather than variable incubation times and hence provide only upper bounds on the times for nucleosome repositioning. More explicitly, *shorter* incubation times would be necessary to test our prediction that repositioning takes place on a time scale as short as seconds. In addition, to test the dependence of rates on DNA/histone binding energies, for example, one would need to carry out experiments for a wide range of concentrations of mono- and di-valent salts. Since the fluctuation energy $\Delta U$ depends essentially linearly on the binding energy $\lambda$, and the rate varies in turn exponentially with $\Delta U$, it is the dependence of $\lambda$ on salt concentrations that will be most important. For example, divalent cations (e.g., Mg$^{2+}$) *enhance* the binding through screening of the self-repulsions of the wrapped DNA, and thereby *suppress* the nucleosome repositioning rate, consistent with observations [6]. On the other hand, divalent cations also compete against proteins for binding to nucleic acids, so they might well *destabilize* the wrapping of DNA on the nucleosome. Since this diffusion rate depends in the usual inverse way on viscosity [Eq. (5)], further checks can come from measurements in which additives are used to significantly increase the effective $\eta$ of the host aqueous solution.

More direct mechanisms for repositioning, which do *not* involve loop formation, can be ruled out by the following arguments. First, one might imagine a *sliding/slipping* motion according to which the chain simply undergoes an overall translation as it slips about the histone octamer. But this would require full unbinding of the chain at each step, and hence as much as *tens* of $k_B T$, intolerably more energy than is thermally available. Second, one might imagine a *rolling* of the histone octamer; fluctuations lead to unbinding of some length of DNA at one end of the nucleosome and simultaneously to wrapping of a comparable length at the other end. While large energies are not required here, the motion is impossible for *geometric* reasons – "rolling" of this kind necessarily involves the chain being progressively displaced from the octamer complex since the histone core is a cylinder of *finite* height (comparable to its diameter).

A final remark concerns connections between the nucleosome diffusion processes discussed here and the motion of an ATP-hydrolyzing motor protein like RNA polymerase. The typical speed *v* of such a protein as it moves along a DNA chain is tens to hundreds of base pairs per second, i.e., about $10^{-6} \, cm/s$. To what extent is it slowed down" by a nucleosome that it encounters? Equivalently, what force does it need to exert on the nucleosome to move it at a speed of $v = 10^{-6} \, cm/s$? The mobility $\mu$ of the nucleosome is given in the usual way by $D/k_B T$, with *D* the diffusion constant calculated earlier. Using our estimate of $10^{-12} \, cm^2/s$ for *D*, it follows that the force $F = v/\mu$ is of the order of a few tenths of a *pN*, i.e., small compared to the maximum force that the motor protein can generate [13]. This suggests that polymerases can move through a nucleosome without being slowed down by any "extra" friction; thermal fluctuations suffice to account for the repositioning of the histone octamer.

*Acknowledgements*: We are grateful to S. Granick for sharing experimental results prior to publication. This work was supported by the National Science Foundation under Grant DMR-9708646.